\pdfoutput=1

\documentclass[letter]{aa} 
\usepackage{natbib}
\usepackage{graphicx}
\usepackage{txfonts}
\newcommand{\dg}{$^\circ$}

\newcommand{\brg}{Br$_{\gamma}$~}
\newcommand{\zcma}{Z~CMa~}

\begin{document}
   \title{The 2008 outburst in the young stellar system Z~CMa
   \thanks{Based on observations collected at the VLTI
   (ESO  Paranal, Chile)  with programs 282.C-5031,
   082.C-0376, 084.C-0162}}
\subtitle{I.  Evidence of an enhanced bipolar wind on the AU-scale}

\author{M.~Benisty\inst{1},   F.~Malbet\inst{2},  C.~Dougados\inst{2},
  A.~Natta\inst{1},     J.B.~Le~Bouquin\inst{2},     F.~Massi\inst{1},
  M.~Bonnefoy\inst{2}, J.~Bouvier\inst{2}, G.~Chauvin\inst{2},  O.~Chesneau\inst{3}, 
  P.J.V.~Garcia\inst{2,4},    K.~Grankin\inst{5},    A.~Isella\inst{6},
  T.~Ratzka\inst{7}, E.~Tatulli\inst{2}, L.~Testi\inst{8},
  G.~Weigelt\inst{9}, E.T.~Whelan\inst{2}} 

\offprints{benisty@arcetri.astro.it}

\institute{INAF-Osservatorio  Astrofisico  di  Arcetri,  Largo
          E.~Fermi 5, 50125 Firenze, Italy
          \and
          Laboratoire d'Astrophysique de  Grenoble, CNRS-UJF UMR 5571,
          414 rue de la Piscine, 38400 St Martin d'H\`eres, France
          \and
          Laboratoire A.~H.~Fizeau, UMR~6525, Universit\'e de Nice-Sophia
          Antipolis, Parc Valrose, 06108 Nice Cedex 02, France
          \and
          Universidade do Porto,  Faculdade de Engenharia, SIM Unidade
          FCT 4006, Rua Dr. Roberto Frias, s/n P-4200-465 Porto, Portugal
          \and
          Crimean  Astrophysical Observatory,  98409  Nauchny, Crimea,
          Ukraine
         \and
          Caltech, MC 249-17, 1200 East California Blvd, Pasadena, CA 91125, USA
          \and
          Universit\"ats-Sternwarte M\"unchen Scheinerstr. 1. 81679 M\"unchen, Germany
          \and
          European  Southern Observatory,  Karl-Schwarzschild-Strasse  2, 85748
          Garching, Germany
          \and
          Max Planck  Institut f\"ur Radioastronomie, Auf dem  H\"ugel 69, 53121
          Bonn, Germany \\}

 \date{Accepted June 26, 2010}

  \abstract
   {Accretion is a fundamental process in star formation. Although the
   time evolution  of accretion  remains a  matter of  debate, 
   observations and modelling studies  suggest that episodic outbursts of
   strong  accretion  may  dominate   the  formation  of  the  central
   protostar.  Observing young  stellar objects  during  these elevated
   accretion  states is crucial  to understanding  the origin  of unsteady
   accretion.} 
   {\zcma  is a  pre-main-sequence  binary system  composed of  an embedded
     Herbig  Be  star, undergoing  photometric  outbursts,  and  a FU~Orionis
     star.  This system therefore  provides a  unique opportunity  to study
     unsteady  accretion  processes.   The  Herbig  Be  component  recently
     underwent its largest optical photometric outburst detected so far. We
     aim  to constrain  the  origin  of this  outburst  by studying  the
     emission region of the HI \brg line, a powerful tracer of
 accretion/ejection processes on the AU-scale in young stars. }
 {Using the AMBER/VLTI instrument at spectral resolutions of 1500 and
   12 000, we performed spatially and spectrally resolved
interferometric observations
   of the hot gas emitting across the \brg emission line, during and
   after the outburst. From  the visibilities and differential phases,
   we   derive  characteristic   sizes  for   the  \brg   emission  and
   spectro-astrometric  measurements  across the  line,  with  respect to  the
   continuum. } 
 {We find  that the line  profile, the
   astrometric signal, and the  visibilities are inconsistent with the
   signature of  either a  Keplerian disk or  infall of  matter.  They
   are, instead, evidence of a bipolar wind, maybe partly seen through a disk
   hole inside the dust sublimation radius.  The disappearance of the \brg emission
line after the outburst suggests  that the outburst is related to a
   period of strong mass loss rather than a change of the extinction along
   the line of sight.  }  
 {Apart from the photometric increase of the system, the main consequence of the
 outburst is  to trigger  a massive bipolar  outflow from the  Herbig Be
 component. Based on these conclusions, we speculate that the origin
 of the outburst is an event of enhanced mass accretion,
 similar to those occuring in EX~Ors and FU~Ors.} 


   \authorrunning{Benisty et al.}
   \titlerunning{Evidence of an enhanced bipolar wind on the AU-scale in Z~CMa}

   \maketitle
%

\section{Introduction}
Accretion plays an important role in star and planet formation. For
many years,  it was considered  to be a slow  quasi-stationary process
\citep[e.g.,][]{stahler98}, occurring mostly through a viscous disk
ending  in  its  inner  part   by  a  boundary  layer  with  the  star
\citep[e.g.,][]{bertout88} or by
magnetospheric funnels \citep[e.g.,][]{konigl91, calvet92}.  
However, this scenario has been challenged by observations
\citep[e.g.,][]{kenyon90,  evans09} that  suggest  that the  accretion
process could be time-variable and occur quickly by means of short high mass
accretion rate bursts.  Studying the very inner region of 
a young stellar object that is known to experience episodic photometric
outbursts  is thus  of  prime  importance to  understand  the role  of
accretion in the formation of the star and its environment. 

\zcma is a pre-main-sequence binary with a separation of 0.1''
\citep{koresko91, barth94} located at a distance estimated from 930 to 1150~pc
\citep[e.g.,][]{claria74, kaltcheva00}.  The primary, embedded in a dust
cocoon,   was    identified   as    a   Herbig~Be   star    based   on
spectropolarimetry \citep{whitney93}. It is surrounded by
an inclined disk, possibly a circumbinary disk, as inferred from millimeter observations
\citep{alonso09},  and  dominates  the  infrared continuum  and  total
luminosity of the system. In contrast, the secondary is the major source of continuum emission at visual wavelengths. Although the secondary has not undergone a large outburst
this century, it  was identified as a FU~Or object  based on its broad
double-peaked optical absorption lines, which are typical of a circumstellar disk that
undergoes a strong accretion, and spectral type of
F-G \citep{hartmann89}.  In the past twenty years, the \zcma 
system  exhibited  repeated   brightness  variations,  of  $\sim$0.5-1
visual magnitude, which were attributed to the Herbig~Be star 
\citep[e.g.,][]{vandenancker04}. 
\zcma is clearly associated with 
a     bipolar    outflow    that     extends    to     3.6~pc    along
PA$\sim$240\dg~\citep{poetzel89, evans94}.  \cite{garcia99} detected a
1"x0.24" 
micro-jet in the [OI]~6300\AA~line in the same direction, and concluded
that the optical emission-line spectrum and the jet are associated with the
primary.  
However, the innermost environments of the \zcma components have been
poorly studied.  Two broad-band interferometric measurements
have been obtained, allowing only characteristic sizes of the
K-band continuum emission to be 
derived  \citep{monnier05,millan06}.    

In January 2008, Z~CMa's brightness increased by about two visual
magnitudes \citep{grankin09}, representing the 
largest   outburst  observed   in  the   past  90   years.   Based  on
spectropolarimetric observations, \cite{szeifert10} concluded that this outburst is associated with
the Herbig~Be star.  

The study  presented in  this paper is  part of a  large observational
campaign targeting Z~CMa during  this outburst that aims to understand
its origin (Bonnefoy~et~al.~in~prep., Bouvier et al.~in prep., Ratzka 
et al.~in prep., Whelan et al.~in prep.).  The 
overall spectral energy distribution  of the system is strongly
modified  during the  outburst at  wavelengths shorter  than 10~$\mu$m
(Bonnefoy et al., in prep.;  Ratzka et al., in prep.), which indicates
that the outburst originates close to the star.  To directly probe the
morphology of  the hot gas in the  inner AUs, we took
advantage of the spatial and spectral resolution available at the VLTI
to perform $\mu$-arcsecond spectro-astrometry. We resolved the K-band
emission of the hot gas surrounding \textit{each} star at the milliarcsecond
resolution.  This paper reports the first spatially \textit{and}
spectrally resolved observations in  Br$_{\gamma}$ of a young star. We
also observed the binary system after the outburst.  In
Sect.~2,  we present  the  observations and  the  data processing.  In
Sects.~3 and 4, we describe and discuss the results.

\section{Observations and data processing}
\zcma was observed at the Very Large Telescope Interferometer
\citep[VLTI;][]{vlti1}, using the AMBER instrument
that allows the simultaneous combination of three
beams in the near-infrared \citep{petrov07}.  
The instrument delivers spectrally dispersed interferometric observables
(visibilities,  closure phases, differential  phases) at
spectral resolutions up to 12~000.  

In the following, we present K-band observations taken in the medium spectral
resolution mode (MR; R$\sim$1500)  with the 8.2~m Unit Telescopes (UTs)
as well as with the 1.8~m Auxiliary Telescopes (ATs), and in
the high spectral resolution mode (HR; R$\sim$12~000) with the ATs.  
The data were obtained  within programs of Guaranteed Time, Director's
Discretionary Time, and Open Time observations.  \zcma was observed with 11 different baselines of 4 VLTI configurations, during 5~nights in December 2008 and one night in
January 2010.  The longest  baseline is $\sim$120~m corresponding to a
maximum angular resolution of 3.7~mas. A summary
of the observations presented in this paper is given in Table~\ref{tab:obs}.  
With the UTs, the observations are coupled with the use of adaptive
optics and the resulting field of view ranges from 50 to 60~mas. This allowed us to spatially resolve the binary and obtain
\textit{separate}  measurements of  the FU~Or  and the  Herbig~Be.  In
contrast, the ATs field-of-view, ranging from 230 to 280~mas, includes
both stars and the interferometric signal results from both emissions.  In
addition to Z~CMa, calibrators (HD45420, HD60742, HD55137, HD55832) were observed to correct for
instrumental 
effects. All  observations   were  performed  using  the 
fringe-tracker FINITO \citep{lebouquin08}.  

The  data  reduction   was  performed  following  standard  procedures
described in \citet{tatulli07} and \citet{chelli09}, using the \texttt{amdlib} package,
release  2.99,  and the  \texttt{yorick}  interface  provided by  the
Jean-Marie                                                     Mariotti
Center\footnotemark{}\footnotetext{$\textrm{http://www.jmmc.fr}$}.
Raw spectral visibilities, differential phases, and closure phases were extracted for
all the  frames of each  observing file.  A  selection of 80\%  of the
highest quality frames was made and 
consecutive observations were merged to enhance
the signal-to-noise ratio. 
The accuracy of the wavelength/velocity calibration is $\sim$50~km/s.
Because the K-band continuum measured by the ATs (due to both
stars)  is  very  resolved  on long  baselines  (V$\sim$0),  the
observations obtained on  the G1-A0 and K0-A0 baselines  could not be
exploited. The absolute value
of  the visibilities  obtained with  the  UT baselines  could not  be
determined due to random vibrations of the telescopes.  However, this
issue  affects all spectral  channels in  the same  way, and  does not
modify our conclusions.

\begin{table}[t]
\centering
\caption{Log    of    the   observations.    R    is   the    spectral
  resolution. 'FUOr+HBe' specifies when the binary is in the field of view.} 
\label{tab:obs}
\begin{tabular}{cccccc}
 \hline
Date & Baseline & Projected & Position & R & \\ 
& & length (m)& angle ($^\circ$)& & \\
 \hline
 05/12/08 & D0-G1 & 69 & 137 & 1500 & FUOr+HBe \\
 07/12/08 & K0-G1 & 89 & 28 & 12000 & FUOr+HBe  \\
 09/12/08 & K0-G1 & 88 & 24& 12000 & FUOr+HBe \\
 \hline 
 15/12/08 & U2-U3 & 44 & 35& 1500 &  \\
 & U3-U4 & 62 & 107 & & \\
 & U2-U4 & 86 & 78 & &  \\
 16/12/08 & U1-U2 & 56& 35& 1500 &  \\
 & U2-U4 & 77& 88& &  \\
 &  U1-U4 &120 & 66& & \\
 \hline
 10/01/10 & D0-G1 & 71 & 133& 1500 & FUOr+HBe \\
 \hline
\end{tabular}
\end{table}


 \begin{figure*}
 \sidecaption
   \includegraphics[width=13cm]{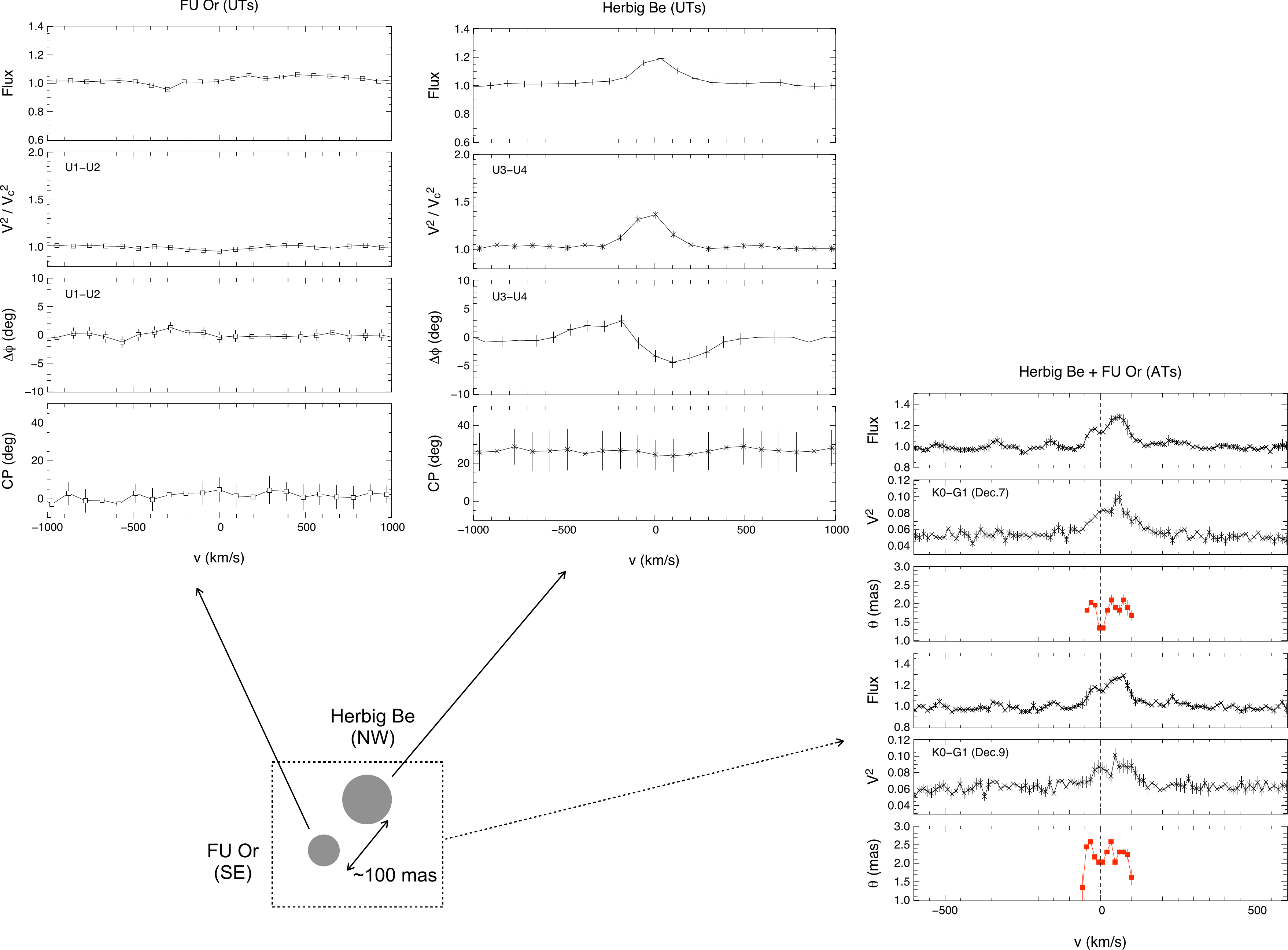} 
      \caption{  Interferometric measurements  of  the Z~CMa
    binary system  in December~2008.  The binary  system is
    sketched at the bottom left. Left panels: 
    MR  spectrum,  squared visibilities  (normalized  to the  continuum
    ones, V$_{\rm{c}}^{2}$), differential 
    phases ($\Delta\Phi$),  and closure phases (CP)  measured for the  FU~Or with
    the  UTs   during  the  outburst.   Middle  panels:   same  for  the
    Herbig~Be. Right panels: ATs observations that include both stellar emissions
    in  its   field  of  view:   HR  spectra,  visibilities,   and  the
    characteristic sizes of the \brg emission (see text for details).} 
     \label{fig:all}
 \end{figure*}


\section{Results}
We recall that the  visibilities provide information about the spatial
extent  of the  emission,  and decrease  as  the extension  increases.
Differential phases provide a measurement of the 
photocenter  displacements across the sky,  projected along  the baseline
direction. They can therefore be converted into differential spectro-astrometric
shifts.  They are measured relative to the continuum, for which we assume a zero phase. 
Finally, the closure phases are related to the asymmetry of the
brightness   distribution  (\textit{e.g.},   they  are   null   for  a
point-symmetric object).  

\begin{figure}[t]
 \centering
  \includegraphics[width=0.33\textwidth]{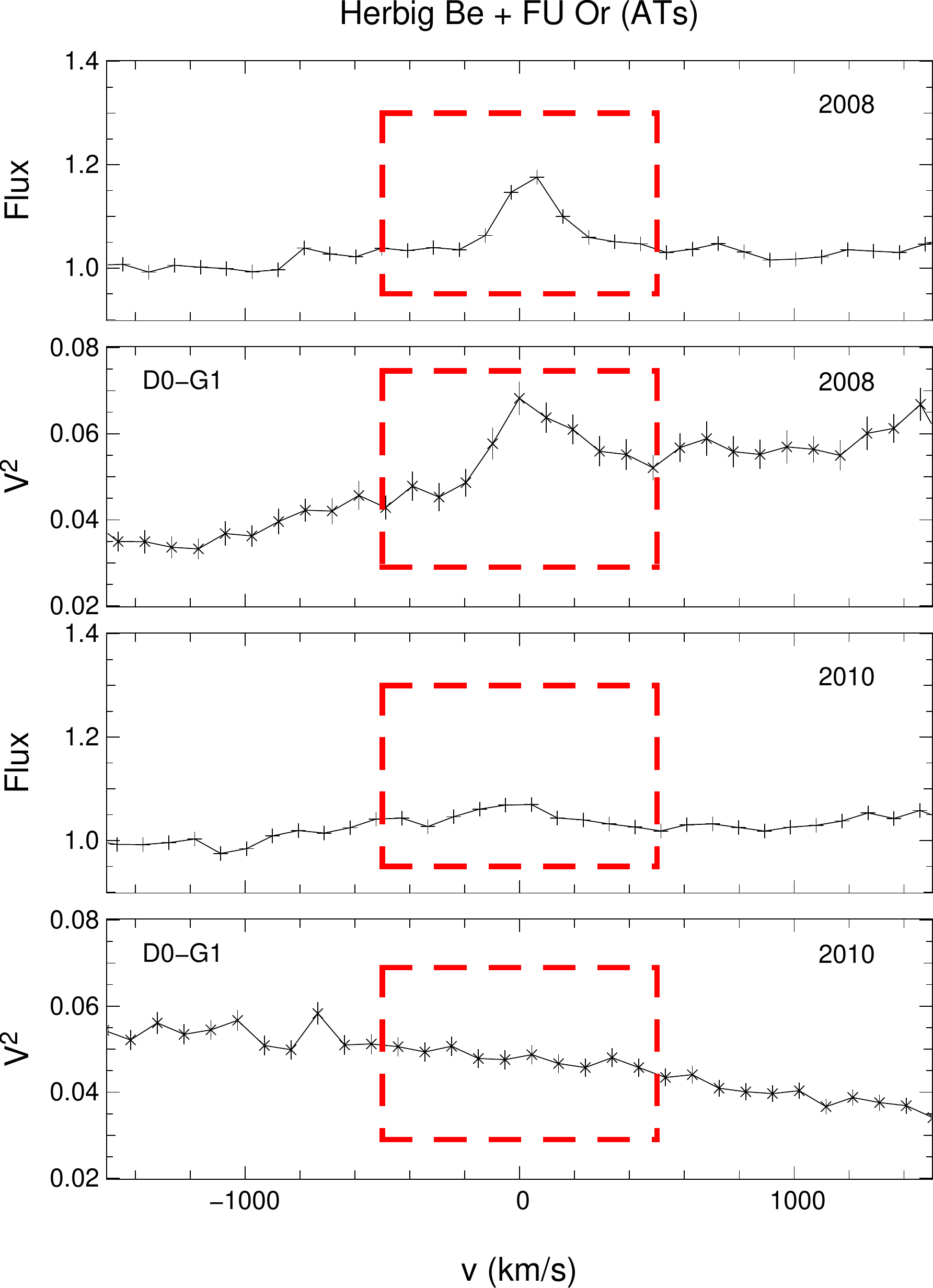}  
  \caption{\label{fig:all2} MR normalized spectra and visibilities during (2008)
  and after the outburst (2010). The slopes of the visibility curves depend on the binary
  characteristics and on the observing set up, that differs in the two
  observations. Note  the disappearance of the \brg  line signature in
  the spectra and the visibility after the outburst (red dashed
squares). } 
\end{figure}

We show  in Figs.~\ref{fig:all} and~\ref{fig:all2}  a subsample of
the observations that illustrate the main characteristics of the data.  Since the absolute 
values  of the  visibilities measured  with  the UTs  are unknown,  we
normalized the continuum values to 1 -- even though the emission
is resolved.  The left and middle columns of Fig.~\ref{fig:all} present
examples of  the MR  observations obtained with  the UTs for  each star
during the  outburst. Each column  includes a spectrum  (normalized to
the continuum), squared visibilities, differential phases, and closure phases.  For
the FU~Or (left panels), within the error bars, the spectrum
shows \brg in neither emission nor absorption. Consequently, no change
in the visibilities or phases across the line is expected/seen.  
In contrast, the Herbig~Be star exhibits a clear \brg line in
emission (middle panels), although at this spectral resolution
($\Delta v\sim$200~km/s),  the line  is not spectrally  resolved.  The
visibility increases through the line and the 
differential phases produce an S-shape variation.  The closure phases 
differ from zero, with values of 25$^\circ 
\pm$12$^\circ$.  The phase, line, and visibility signals are present 
from $\sim$-600 to 500~km/s, although because of the low line-to-continuum
ratio in the extended wings, the flux and visibilities appear narrower.
Within  the  large errors,  no  variation  in  the closure  phases  is
detected across the line. 

The right  part  of the  figure shows  measurements
obtained with the ATs, 
\textit{i.e.}, with both stars in the field of view. In this case, the level of
continuum is determined by both stellar components. 
These panels present observations obtained in HR. 
In this case,  the line is spatially and  spectrally resolved ($\Delta
v\sim$25~km/s), and the spectra exhibit a 
clear  double-peaked and  asymmetric  profiles, with  less emission  at
blueshifted velocities.  The spectral visibilities present a similar profile.  

Finally, Fig.~\ref{fig:all2} compares the spectra and the
visibilities obtained during and after the outburst: the emission
line,  and the  signature  in the  visibilities,  disappear after  the
outburst. Plotted within a large velocity range, the visibilities show
a typical signature of  binarity (\textit{i.e.}, a cosine modulation),
in agreement with the system main characteristics 
(separation, position angle, flux ratio; Bonnefoy et al., in prep.). 




From the visibilities, one can locate the emission at
each  velocity and  distinguish between  various scenarios  capable of
producing the line.  The visibility increase within 
the line implies that the  \brg emitting region  is more compact
than the one responsible for the continuum.  To derive the characteristic sizes
of the region emitting \brg only, for each spectral channel of the
HR measurements, one 
has  to substract  the  underlying continuum  to  first determine  the
visibility of  the line  only \citep{weigelt07}.  These  estimates can
only be performed using the data gathered with the ATs, for which reliable absolute values for the \brg
visibilities are obtained.  Using a model of an uniform ring,
the emission in the line has a typical extension (ring diameter) of $\sim$1.6~mas at zero 
velocity,  and $\sim$2.5~mas  at  higher velocities  ($\sim$100~km/s),
\textit{i.e.}, from $\sim$1.5 to $\sim$2.6~AU, 
depending on the distance.  As the continuum emission measured with the ATs
includes  both  stars,  it  is  not  direct to  establish
the typical size of the Herbig~Be continuum. In contrast, the UTs data
include only one stellar component. Although no 
absolute visibility values can be obtained, size ratios between the 
line and the continuum can be derived.  Using the sizes
previously estimated  for the line from  the ATs data,  typical sizes of
$\sim$3.4~mas ($\sim$3.6~AU) for  the Herbig~Be K-band continuum can
be determined, in agreement  with the previous estimate \citep[$\sim$3.9~mas
in 2004;][]{monnier05}. Considering a dust sublimation temperature
around  1500-2000~K  \citep{pollack94},  and  the  stellar  properties
determined by \cite{vandenancker04}, the inner edge of the dusty disk must be 
located at $\sim$4-7~AU, in  agreement with our findings. An asymmetry
in the inclined  inner disk could explain the  non-zero closure phases
measured at a level similar to other 
 Herbig~AeBe stars \citep{kraus09, benisty10}.  

The differential phases $\Delta\Phi$ can be expressed in terms of
photocenter    displacements    $p$    (in   arcseconds),    following
\citet{lachaume03}, given by $p=-2\pi\Delta\Phi/B\lambda$, 
where $\lambda$ and $B$ are  the  wavelength and the projected
baseline length of the
observations, respectively.  $p$ is the projection  along the baseline
direction, of the 2D photocenter vector $\vec{p}$ in the plane of the sky
(\textit{i.e.}, of a spectro-astrometric signal). We fitted all the differential phases along the 6 available baselines
with  a  single  vector  $\vec{p}$, independently  of  each  spectral
channel.  The  results are presented in  Fig.~\ref{fig:disp}. The left
panels show the differential phases and the best solution
for $\vec{p}$. The middle plot gives $\vec{p}$ in a 2D map of the plane of the
sky.        Clear      asymmetric      displacements,       up      to
$\sim$150~$\mu$-arcseconds, are observed, both at red-shifted
and blue-shifted velocities. 
In this case,  $\vec{p}$ accounts  for the  emission of
both  the   line  and   the  continuum.  Substracting   the  continuum
contribution    to    determine    the   photocenter    displacements,
$\vec{p}_{\rm{Br}_\gamma}$, due to the line only, is difficult, as it
has to be done in the complex visibility plane. We
provide  such an  attempt  in the  velocity  range where  the line  is
clearly detected ([-350;350]~km/s, with line-to-continuum ratio larger
than  1.05).  As  can  be seen  in Fig.~\ref{fig:disp},  right, the
displacements  are much larger  (up to  $\sim$1~mas) with  the largest
measured    at    the   highest    velocities,    and   appear    more
spread. Nonetheless, the observed asymmetry is 
still  consistent with  the closure  phase measurements  that  show no
change through the line, within the large errors. 

\section{Evidence of a bipolar wind} 
As has already been discussed in previous studies \citep{kraus08,
eisner09},  the  \brg  line  could   be  emitted  by a  variety  of
  mechanisms, such as accretion of matter onto the star, in
  a gaseous  disk, or in  outflowing matter.  The spectra  obtained at
  high spectral resolution show a double-peaked 
and asymmetric profile  that can be interpreted in  the context of the
formation  of optically  thick lines  in  a dense  environment with  a
temperature gradient \citep{cesa95, kurosawa06}. 

\begin{figure*}[t]
 \centering
\begin{tabular}{ccc}
  \includegraphics[width=0.3\textwidth]{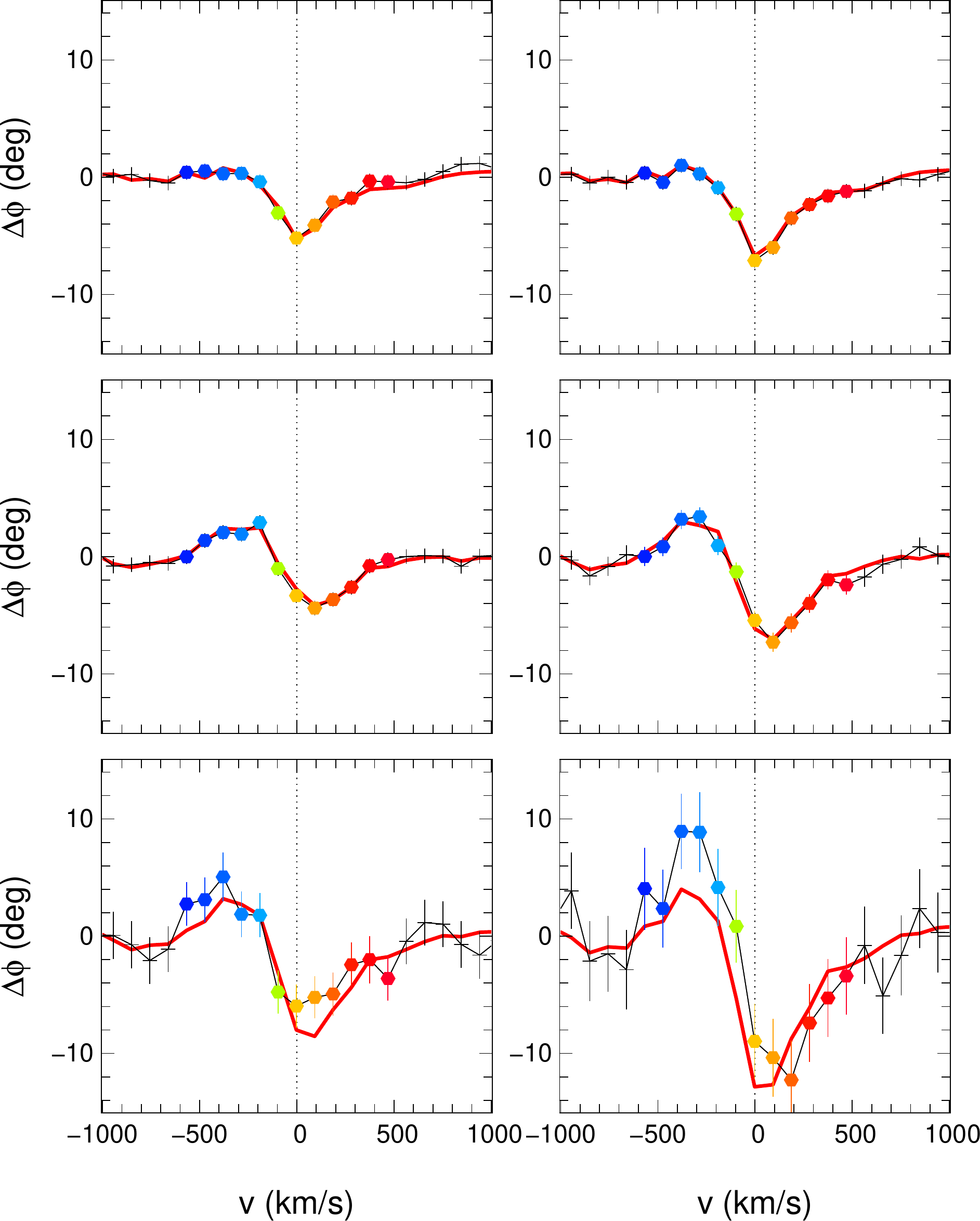}&
  \includegraphics[width=0.34\textwidth]{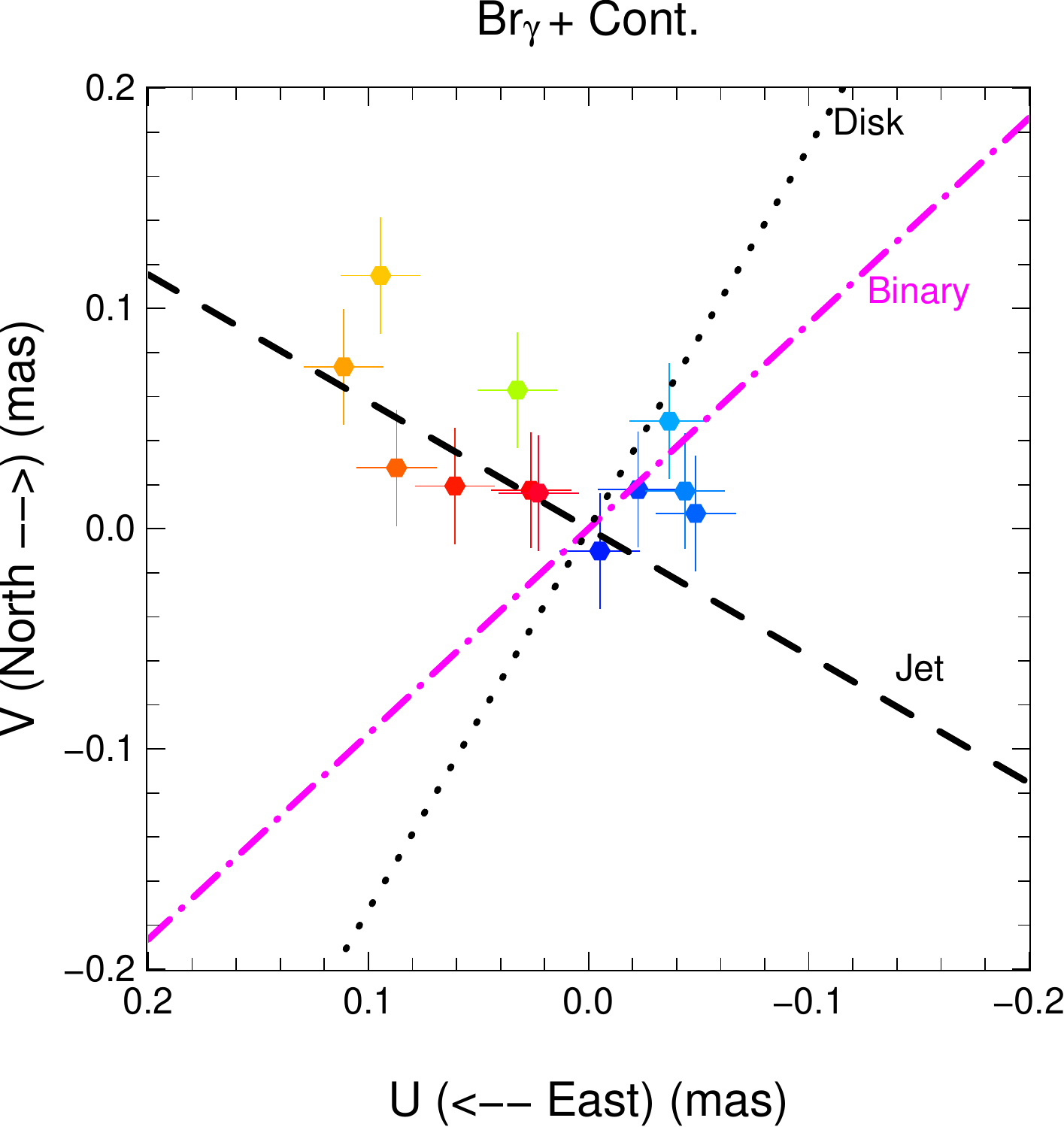}   &
  \includegraphics[width=0.34\textwidth]{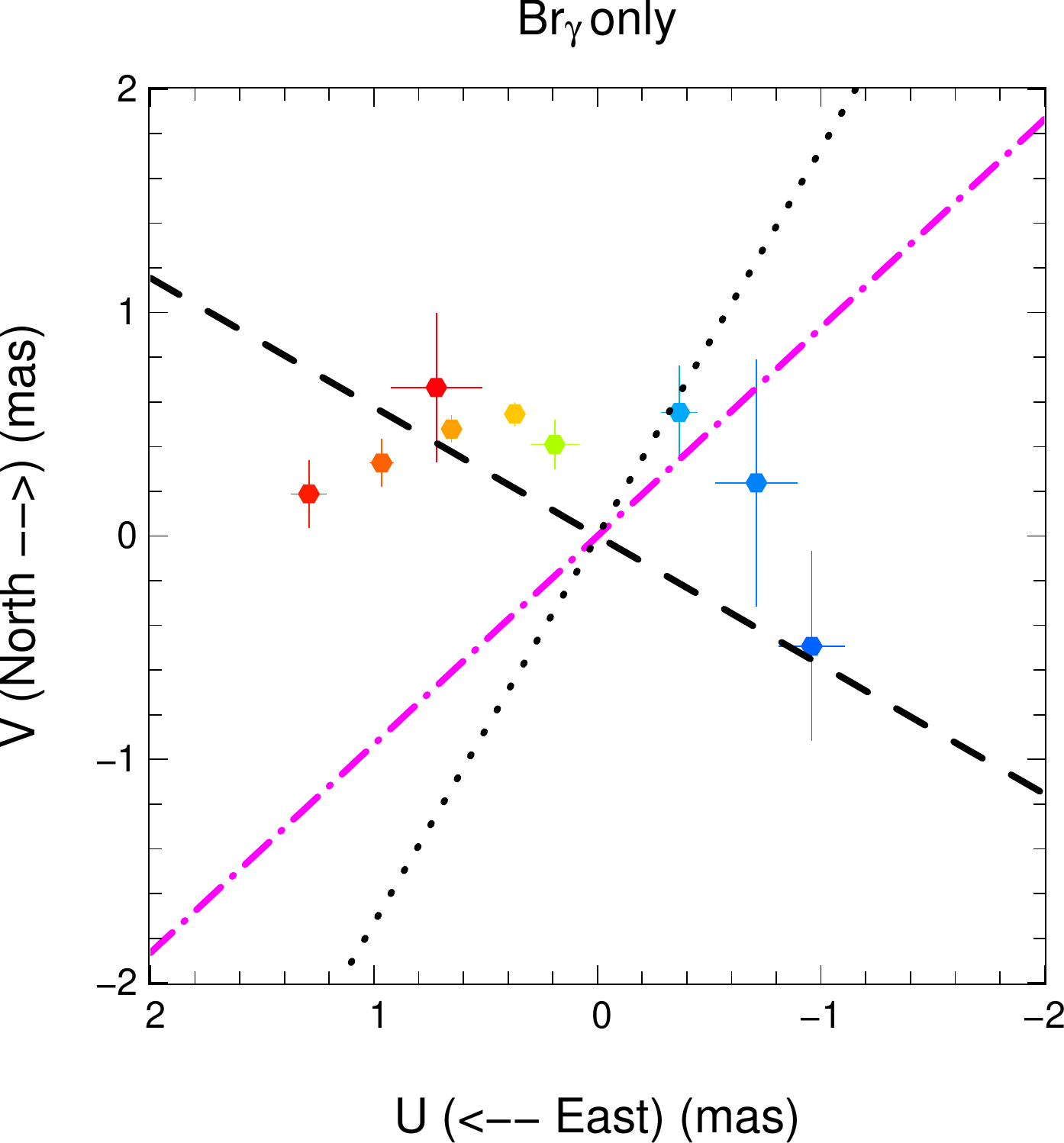}
\end{tabular}
 \caption{\label{fig:disp} Left: differential phases measured with the
  UTs (black  crosses and lines).   The dots in  different colors
    represent   various  velocity   channels,  from   dark   blue  (for
    v$\sim$-600~km/s) to  dark red (for  v$\sim$+500~km/s).  The red
    line is the 2D astrometric solution $\vec{p}$($\lambda$). 
    Middle: 2D representation of $\vec{p}$($\lambda$)
    within -600 and 500 km/s. The different colors code the velocity channels as
    represented in the left panel. The position angles of the binary
    (dashed-dotted line), of  the large-scale jet (dashed  line) and the
    direction perpendicular to the jet (dotted line) are overplotted.  Right: same for
    $\vec{p}_{\rm{Br}_\gamma}$($\lambda$),  after  substraction  of  the
    continuum contribution. This was done within a narrower interval
    (-350 to 350~km/s) because of the low line-to-continuum ratio in the wings.} 
\end{figure*}

Formation of  the \brg  line  in an
infalling envelope of gas can be ruled out. Considering that the line
excitation  temperature increases  towards the  star, if  the  line is
emitted in infall of matter,  or accretion flows, the profile would be
double  peaked but  with an  opposite  asymmetry to  what is  observed
\citep[Fig.~\ref{fig:all}, right; \textit{i.e.},  with a lower emission at redshifted velocities;
see][]{hartmann94, walker94}. In addition, in that case, the smallest 
extension  and  photocenter displacements  would  be  expected at  the
highest velocities,  which is in  disagreement with our  findings (see
Figs.~\ref{fig:all} and \ref{fig:disp}).

The possibility that the \brg line forms in the hot layers of the
gaseous disk can also be ruled out.  If one considers that the
circumstellar disk surrounding the Herbig~Be 
star is perpendicular to the large-scale jet at PA$\sim$240\dg, it would be
expected that the  velocities projected onto the line  of sight cancel
out  along  the   semi-minor  axis,  while  large  spectro-astrometric
displacements   are  seen   along   this  axis   at  high   velocities
(Fig.~\ref{fig:disp}).  Apart from this, the displacements increase with
velocity while  Keplerian rotation should behave in  the opposite way,
and a phase signal is measured up to high 
velocities ($\sim$500~km/s), which are much larger than the expected Keplerian
velocities ($\sim$100-120~km/s at $\sim$1~AU). It therefore seems 
unlikely that the \brg line is emitted in the disk. 

We consider that the most likely origin of the \brg emission
is a wind.  Strong winds are expected to take place in massive
Herbig~Be \citep{nisini95, malbet07}, and could be responsible for the
\brg emission.  The double peaked and asymmetric line profile is
consistent  with   outflowing  matter  emitting   in  optically  thick
lines \citep[][]{hartmann90}.  The visibilities and the 2D maps of the
astrometric signal also support this 
conclusion as the blue- and  redshifted emissions are located on each
side of the possible disk position angle, with the largest \brg displacements and
characteristic sizes being derived at higher velocities. Our
observations  suggest that  the disk is  slightly  inclined, to
allow both red- and blueshifted 
emissions to be seen.   We may be
seeing the  emission from  a wind partly  through an  optically thin
inner hole in the optically thick dusty disk \citep{takami01,
 emma04}. Alternatively, if the inner gaseous disk were optically thick, we
  may be seeing the redshifted emission through a much smaller hole and via
scattering  on  the disk  surface.  Whether  the  innermost disk  is
  optically thick or  not cannot be determined  with our observations
  and no reliable estimate of  the mass accretion rate exists for such
  high  mass young  stars. However,  the presence  of the  CO overtone
  lines in emission (Bonnefoy et al., in prep) is indicative of a much lower
  mass accretion rate 
  than     those     derived    for     FU~Ors     \citep[$\leq
  10^{-5}$M$_{\odot}.$yr$^{-1}$;][]{calvet91, carr89}.   At the spatial resolutions provided by the VLTI, we
  trace the regions close to the inner disk hole and it is therefore unsurprising that
  we could  detect redshifted emission, while on  scales of 10-100~AU,
  the redshifted lobe is obscured by the circumstellar disk \citep[Whelan et al., in
  prep.]{poetzel89,   garcia99}.   During   this  outburst,   deep
blueshifted absorption was detected 
  in  the  Balmer  lines  from  zero velocity  to  $\sim$700~km/s,  in
  addition to the absence of redshifted emission at similar velocities 
\citep[Bouvier~et~al.,~in~prep.]{szeifert10}, 
supporting our conclusion that there is a strong wind in the
Herbig~Be. \\ 
\indent Could our new observations be tracing the inner parts of the parsec scale outflow? As shown
in Fig.~\ref{fig:disp}, the astrometric signal is 
detected at a slightly different position angle 
and at these spatial scales, it is unlikely that the jet is already
collimated.  Our observations exclude  a fully spherical wind since in
that case no displacement would be expected between the redshifted and
blueshifted emission lobes. The derived spectro-astrometric signatures
favor  a  bipolar  wind, maybe  unrelated  to  the  jet, but  can  not
determine whether  its geometry  is that of  a disk-wind or  a stellar
wind.  

After  detecting  the  same  level  of optical  polarisation  in  both
continuum  and   spectral  lines   along  a  position   angle  roughly
perpendicular to the large-scale jet, \cite{szeifert10} concluded that
this outburst is related to a  change in the path along which the photons
escape from the dust cocoon. 
The disappearance of the \brg emission line, with respect to the
continuum, after the outburst, suggests that its emission is related to
the outburst.  A strong mass ejection event could account for 
the deep blueshifted absorption features seen in the Balmer lines that
are emitted close to the star as well as for the
\brg line emitted in outer layers of the wind.  Outside the
outburst, the wind disappears or is more likely to be maintained at a much
smaller mass loss rate.  Based on these 
conclusions, one can speculate about the origin of the outburst, as
being driven by an event of enhanced mass accretion, similar to the EX~Ors
and FU~Ors outbursts \citep{zhu10}. In that case, this would
suggest  a strong link between mass accretion and ejection during the
outburst, probably  coupled with a  magnetic field as  in lower-mass
young stars. 

\section{Conclusions}
We have presented spatially and spectrally resolved
interferometric  observations  of the  K-band  emission  in the  Z~CMa
system.  These observations were performed during the largest
photometric outburst  detected so far,  that occurred in  the innermost
regions of the Herbig~Be star. 

We found that  the \brg line profile, the  astrometric signal, and the
characteristic sizes across the line are inconsistent with a 
Keplerian disk or with infall  of matter.  They are, instead, evidence
of a bipolar wind seen through a disk hole, inside the 
dust  sublimation radius.  The disappearance  of the  \brg emission
line after  the outburst  suggests that the  outburst is related  to a
period  of strong  mass loss.   Based  on these  conclusions, we  have
speculated that  the origin  of the outburst  is an event  of enhanced
mass accretion, and that it does not result from a change in the system
obscuration by dust. If this were valid, our results would suggest that
the link between 
mass accretion  and ejection as  observed for quiescent T  Tauri stars
can also be at play in more massive young stars, and in high-accretion
states. 
  
Finally, this paper illustrates the great potential of the combination
of  spectro-astrometric and  interferometric techniques  for observing
structures on $\mu$-arcsecond scales.




\begin{acknowledgements}
 We  thank  the  VLTI  team  at  Paranal,  as  well  as  R.  Cesaroni,
 S. Antoniucci, L.  Podio, P.~Stee and M.~van den  Ancker for fruitful
 discussions.    We   thank   the   anonymous  referee   for   helpful
 comments.  M.B.  acknowledges funding  from INAF  (grant ASI-INAF
 I/016/07/0).  
\end{acknowledgements}

\bibliographystyle{aa}
\bibliography{14776}

\end{document}